\documentstyle[epsfig]{mn}
\begin{document}

\title[Galactic centre LMXBs]
{Simultaneous radio and X-ray observations of Galactic Centre low mass
X-ray binaries}
\author[Stephan G. Berendsen et al.]
{
Stephan G.H. Berendsen$^{1,2}$, Robert Fender$^1$, Erik Kuulkers$^{3,4}$, 
J. Heise$^3$,\cr M. van der Klis$^1$\\
$^1$ Astronomical Institute `Anton Pannekoek', University of Amsterdam,
and Center for High Energy Astrophysics, Kruislaan 403, \\
1098 SJ, Amsterdam, The Netherlands\\
$^2$ F.E.O. Hogeschool v. Utrecht, Archimedeslaan 16, 3584 BA Utrecht,
The Netherlands\\
$^3$ Space Research Organization Netherlands, Sorbonnelaan 2, 
3584 CA Utrecht, The Netherlands \\
$^4$ Astronomical Institute, Utrecht University,
P.O.\ Box 80000, 3507 TA Utrecht, The Netherlands\\
}

\maketitle

\begin{abstract}

We have performed simultaneous X-ray and radio observations of
thirteen Galactic centre low-mass X-ray binaries in 1998 April using
the Wide-Field Cameras onboard BeppoSAX and the Australia Telescope
Compact Array, the latter simultaneously at 4.8 and 8.64 GHz.  We
detect two Z sources, GX 17+2 and GX 5-1, and the unusual `hybrid'
source GX 13+1.  Upper limits, which are significantly deeper than
previous non-detections, are placed on the radio emission from two
more Z sources and seven atoll sources. Hardness-Intensity diagrams
constructed from the Wide-Field Camera data reveal GX 17+2 and GX 5-1
to have been on the lower part of the horizontal branch and/or the
upper part of the normal branch, at the time of the observations, and
the two non-detected Z sources, GX 340+0 and GX 349+2, to have been on
the lower part of the normal branch. This is consistent with the
previous empirically-determined relation between radio and X-ray
emission from Z sources, in which radio emission is strongest on the
Horizontal branch and weakest on the Flaring branch.  For the first
time we have information on the X-ray state of atoll sources, which
are clearly radio-quiet relative to the Z sources, during periods of
observed radio upper limits. We place limits on the linear
polarisation from the three detected sources, and use accurate radio
astrometry of GX 17+2 to confirm that it is probably not associated
with the optical star NP Ser. Additionally we place strong upper
limits on the radio emission from the X-ray binary 2S 0921-630,
disagreeing with suggestions that it is a Z-source viewed edge-on.

\end{abstract}
\begin{keywords}

binaries: close -- radio continuum : stars -- X-rays : stars q--
polarization -- astrometry

\end{keywords}


\section{Introduction}

The Z and atoll sources are thought to be X-ray binaries containing
low-magnetic-field ($10^8$--$10^9$ Gauss)
neutron stars (as many of the atoll, and least one
of the Z sources, show X-ray bursts) accreting from a low mass ($M
\leq $M$_{\odot}$) companion star.  Use of X-ray `colour-colour'
and/or `hardness-intensity' diagrams (CCDs/HIDs) and high-sensitivity
X-ray timing have provided a wealth of information on the accretion
processes in these sources.  Together, the two groups encompass all
the brightest low-mass X-ray binaries, and they are clustered in the
direction of the galactic bulge. See e.g. van der Klis (1995, 1999)
for reviews of their X-ray properties.  All the Z sources, and at
least one of the atoll-sources, have been detected at radio
wavelengths (see e.g. Penninx 1989; Hjellming \& Han 1995; Fender
\& Hendry 2000). As with all X-ray binaries detected
at radio wavelengths, high brightness temperatures appear to preclude
thermal emission processes, and synchrotron emission is the favoured
emission mechanism. However, other measurements, notably detection of
linear polarisation, are required to confirm this interpretation.  In
several cases high resolution radio observations have resolved the
radio emission into jet-like structures, sometimes with components
moving at relativistic velocities (e.g. Hjellming \& Han 1995; Fender,
Bell Burnell \& Waltman 1997; Mirabel \& Rodr\'\i guez 1999).

The radio emission from the atoll and Z sources shares some common
properties with that of the persistent black hole X-ray binaries
(Fender \& Hendry 2000) which suggests that their low intrinsic
magnetic fields (believed to be $\leq 10^{11}$ G) have little effect
on the accretion/acceleration process (at least for values of the
accretion rate near to the Eddington limit). The major influence on
the acceleration of radio-emitting relativistic electrons must be the
state of the accretion disc, itself determined by the accretion
rate. Thus at all times we should expect to see some relation between
behaviour in X-ray timing and spectra and radio timing and spectra.

Apart from the unusual source GX 13+1 (and probably also GX 354+0 --
Mart\'\i{} et al. 1998), no atoll source has been reliably detected at more
than one epoch. Previous upper limits can be found in Grindlay \&
Seaquist (1986), Nelson \& Spencer (1988), Cooke \& Ponman (1991); see
Hjellming \& Han (1995) for more details.

\section{Observations}

On 1998 April 1 and 4 we observed several low-mass X-ray binaries in the
direction of the Galactic Centre with the Australia Telescope Compact
Array (ATCA) contemporaneously with X-ray observations of the region with
the SAX Wide Field Camera (WFC). In addition we have used XTE All Sky
Monitor (ASM) data to judge the longer-term trend in the X-ray
behaviour of the sources around the times of our observations.

\subsection{ATCA}

\begin{table*}
\begin{tabular}{cccccccccc}
     &        &   \multicolumn{4}{c}{All baselines} &  \multicolumn{4}{c}{Only baselines $\geq 3$km}\\ 
             && \multicolumn{2}{c}{Total intensity(mJy)} & \multicolumn{2}{c}{Fractional polarisation} & \multicolumn{2}{c}{Total intensity(mJy)} & \multicolumn{2}{c}{Fractional polarisation}\\
Date & Source & 6.3 cm & 3.5 cm & 6.3 cm & 3.5 cm & 6.3 cm & 3.5 cm & 6.3 cm & 3.5 cm \\
\hline
1998 April 1 & 2S 0921-630 & $<0.10$ & $<0.10$ & -- & -- & -- & -- & -- & -- \\
(MJD 50904) & 4U 1820-30 & $<0.14$ & $<0.19$ & -- & -- & -- & -- & -- & -- \\
            & GX 17+2 & $0.6 \pm 0.04$ & $1.3 \pm 0.08$ & $<0.16$ & $<0.14$ & $0.4 \pm 0.04$ & $0.9 \pm 0.03$ & $<0.28$ & $<0.13$ \\
            & GX 3+1 & $<0.18$ & $<0.17$ & -- & -- & $<0.09$ & $<0.15$ & -- & -- \\
            & GX 9+1 & $<0.16$ & $<0.18$ & -- & -- & $<0.12$ & $<0.10$ & -- & -- \\
            & GX 9+9 & $<0.21$ & $<0.21$ & -- & -- & $<0.21$ & $<0.15$ & -- & -- \\
            & GX 5-1 & $<0.23$ & $1.2 \pm 0.35$ & -- &$<0.68$ & $0.4 \pm 0.04$ & $0.6 \pm 0.05$ & $<0.33$ & $<0.23$ \\
            & GX 13+1 & $<0.80$ & $<0.68$  & -- & -- & $1.5 \pm 0.05$ & $1.2 \pm 0.05$ & $<0.08$ & $<0.12$ \\
\hline
1998 April 4 & 4U 1636-53 & $<0.12$ & $<0.15$ & -- & -- & $<0.17$ & $<0.20$ & -- & -- \\
(MJD 50907) & GX 340+0 & $<1.3$ & $<0.20$ & -- & -- & $<0.30$ & $<0.20$ & -- & -- \\
& GX 339-4 & $<0.11$ & $<0.13$ & -- & -- & $<0.18$ & $<0.22$ & -- & -- \\
& 4U 1705-44 & $<0.12$ & $<0.12$ & -- & -- & $<0.16$ & $<0.20$ & -- & -- \\
& 4U 1735-44 & $<0.13$ & $<0.14$ & -- & -- & $<0.16$ & $<0.20$ & -- & -- \\
& GX 349+2 & $<0.11$ & $<0.39$ & -- & -- & $<0.11$ & $<0.16$ & -- & -- \\
\hline  
\end{tabular}
\caption{Total intensity and the fractional polarisation of the 14
sources both with all the baselines and only baselines $>3$km.}
\end{table*}

\begin{figure}
\leavevmode\epsfig{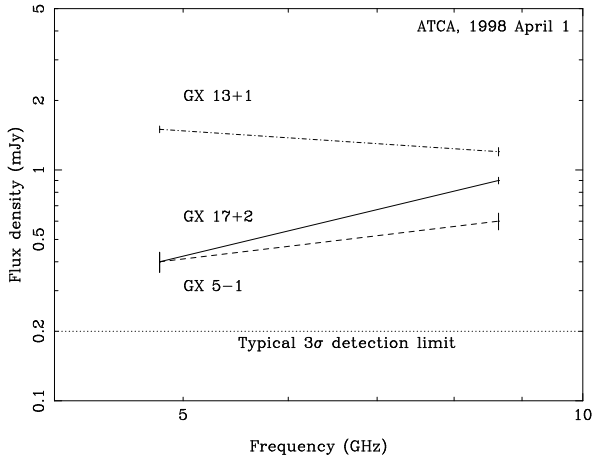}
\caption[]
{4 - 9 GHz radio spectra of the three sources detected, using flux
densities measured from point-source fits to long-baseline ($\geq 3$
km) data only (see text).}
\end{figure}

\begin{figure*}
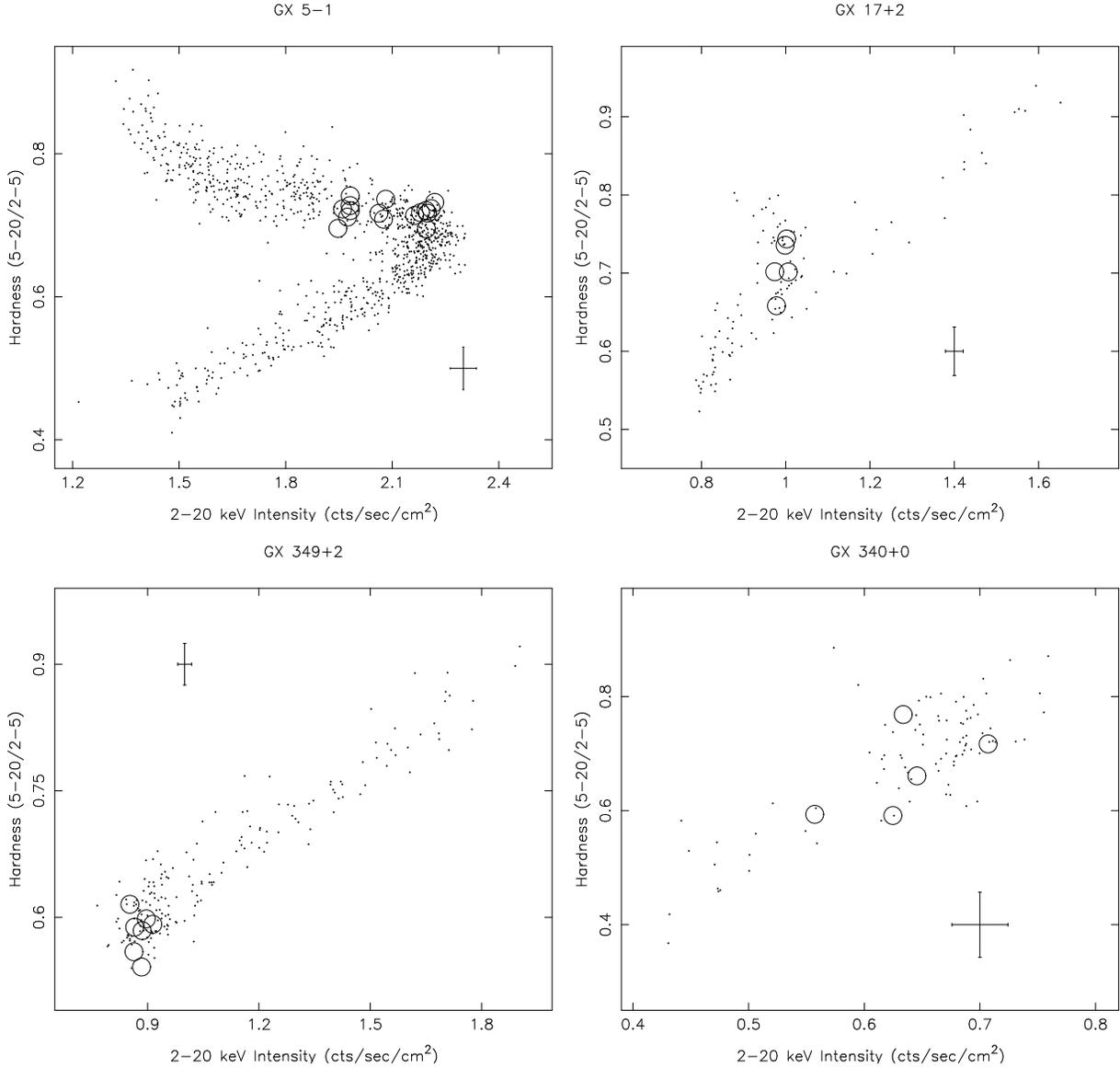

\centerline{\epsfig{file=gx5-1.ps,width=7.5cm,angle=-90,clip}\quad\epsfig{file=gx17+2.ps,width=7.5cm,angle=-90,clip}}
\vspace*{3mm}
\centerline{\epsfig{file=gx349+2.ps,width=7.5cm,angle=-90,clip}\quad\epsfig{file=gx340+0.ps,width=7.5cm,angle=-90,clip}}
\vspace*{1mm}
\caption[] 
{Hardness-Intensity plot of the SAX WFC data for the Z sources during
the period of the ATCA radio observations. All data from the 1998
April 1--5 SAX WFC observations are plotted; open circles indicates
those observations strictly simultaneous with the radio
observations. Typical uncertainties are indicated on each plot. The
integration times are 2 min, 10 min, 16 min and 20 min, for GX 5-1, GX
349+2, GX 17+2 and GX 340+0, respectively.}
\end{figure*}

\begin{figure*}
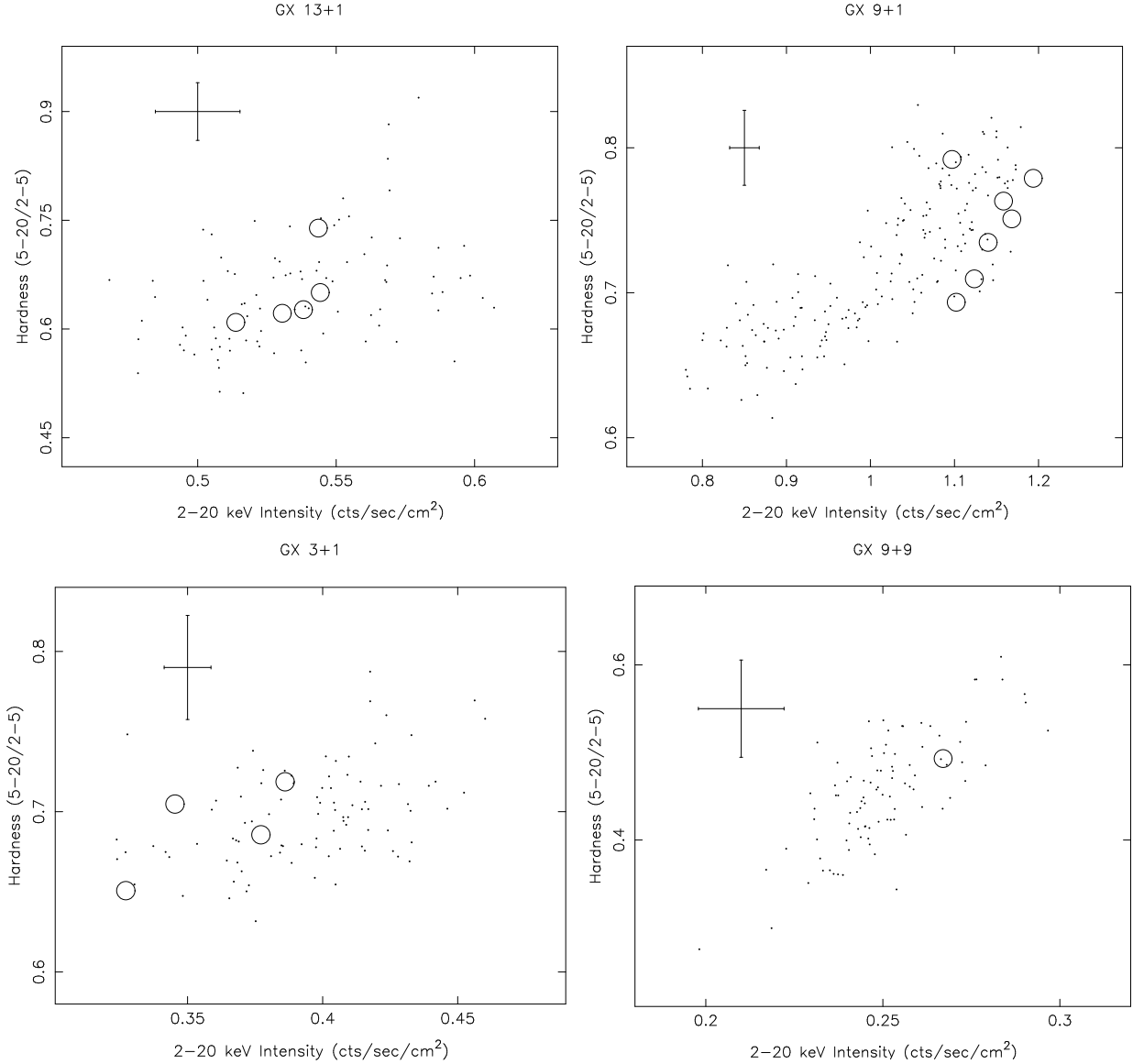

\centerline{\epsfig{file=gx13+1.ps,width=7.5cm,angle=-90,clip}\quad\epsfig{file=gx9+1.ps,width=7.5cm,angle=-90,clip}}
\vspace*{3mm}
\centerline{\epsfig{file=gx3+1.ps,width=7.5cm,angle=-90,clip}\quad\epsfig{file=gx9+9.ps,width=7.5cm,angle=-90,clip}}
\vspace*{1mm}
\
\caption[] { As Fig 2, but for the four atoll sources which were
strongest during the contemporaneous SAX : ATCA observations.  The
integration times are 10 min for GX 9+1 and 20 min for GX 13+1, GX 3+1
and GX 9+9.  }
\end{figure*}

On 1998 1 and 4 April we observed the Galactic Centre X-ray binaries
sources with ATCA, in a high-resolution 6 km configuration. The
observations were made simultaneously at 6.3 cm (4800 MHz) and 3.5 cm
(8640 MHz).  Observations were from 13:00 -- 02:00 UT and 11:00 --
02:00 UT on April 1-2 and 4-5 respectively.  Primary flux
calibration was achieved using PKS 1934-638. Observations consisted of
15-minute snapshots of the targets interleaved with 5 min observations
of a phase calibrator (VLA calibrator B1748-253/J1751-253 and ATCA
calibrator 1646-50 on April 1 and 4 respectively). This resulted in
four or five 15-min observations of each source over the observing
run. The MIRIAD software package (Sault, Teuben \& Wright 1995) was
used to reduce the data.  Flux densities and upper limits were
measured from maps formed by combining all the snapshots for each
source; for those targets clearly detected, point-source fits in the
image plane were used to estimate the flux density.

In addition, on April 1 we observed the low-mass X-ray binary 2S
0921-630 between 06:15 -- 13:00 UT. As with the snapshots of the
Galactic Centre sources, observations involved 15 of every 20 minutes
on the target, with 5 min on the phase calibrator (ATCA calibrator
0843-54). Primary flux calibration was achieved using PKS 0823-500.

Table 1 lists the targets and summarises the detections and upper
limits at 6.3 and 3.5 cm. We list measurements for maps made using all
baselines and also for those made using only baselines $\geq 3$km
(47000 k$\lambda$ and 85000 k$\lambda$ at 6.3 and 3.5 cm
respectively). By selecting only long baselines, we avoid problems
with the sidelobes of emission of very extended objects typically
found near the Galactic Plane. This emission is problematic for
observations of this kind as the emission is very poorly imaged (see
discussion in Burgess \& Hunstead 1995).  In nearly all cases the flux
density measured on the shortest ($\leq$ few 100m) baselines was much
greater than the flux density we were trying to measure for the
point-source target.  In particular, the sources GX 5-1 and GX 13+1
were clearly detected at both frequencies only when using the longer
baselines (GX 13+1 is in a field of bright, extended emission which
was measured at $\sim 0.8$ Jy at 6.3 cm on the shortest baselines and
is not well imaged by five 15-min snapshots).  In almost half of all
the maps made there were confusing sources which were brighter than
the target.  Fig 1 plots the radio spectra of the three sources which
were unambiguously detected at both frequencies.

For the three sources detected, GX 17+2, GX 5-1 and GX 13+1, we were
additionally able to place limits on the linear polarisation of the
emission of a few tens of per cent (see Table 1).

\subsection{SAX WFC}

The Wide Field Cameras (WFCs, Jager et al.\ 1997) onboard the {\it
BeppoSAX} satellite (Boella et al.\ 1997) are identical coded
mask telescopes pointing in different directions. Each camera has a
40$^{\circ}\times 40^{\circ}$ field of view and covers the 2--27\,keV
energy range.  The imaging capability (angular resolution is about 5')
and the good sensitivity ($\sim$5--10\,mCrab in 10$^4$\,s)
allow an accurate monitoring of complex sky regions, like the Galactic
bulge. The BeppoSAX WFCs are carrying out a program of monitoring observations
of the field around the galactic center.  The purpose is to detect
X-ray transient activity, particularly from low-mass X-ray binaries
(LMXBs) whose Galactic population exhibits a strong concentration in
this field, and to monitor the behavior of persistently bright X-ray
sources (see e.g.\ Heise 1998). This program consists of campaigns
during the (northern) spring and autumn of each year. Each campaign
lasts about two months and typically comprises weekly observations.

In this paper we report on part of the X-ray monitoring program during
spring 1998, i.e.\ April 1--5. These observations were performed with
WFC unit 1. In order to investigate the state of the X-ray sources we
observed in the radio, we created hardness vs.\ intensity diagrams,
where hardness is defined as the ratio of the count rates in the
5--20\,keV and 2--5\,keV band, and intensity as the total count rate
per cm$^2$ in the 2--20\,keV band. All X-ray measurements are
corrected for dead-time and background.  Depending on the brightness
of the source we determined average hardness and intensity values
ranging from 2\,min to 40\,min.

Some hardness-intensity diagrams for the brightest of the Galactic
Centre X-ray binaries at the time of our observations are shown in
Figs 2 (Z sources) and 3 (atoll sources).

\begin{table*}
\begin{tabular}{cccccc}
Source & Class & XTE ASM & SAX WFC & Radio ? & Comments \\
       &       & (ct/sec)& (state$^1$) &         &          \\
\hline
GX 17+2 & Z & 41 & Upper NB/Lower HB (see Fig 2) & Yes & \\
GX 5-1 & Z & 65 & Upper NB/Lower HB (Fig 2) & Yes & \\
GX 340+0 & Z & 30 & Upper NB (Fig 2) & No & Previously reliably detected \\
GX 349+2 & Z & 53 & Lower NB (Fig 2)& No & Previously reliably detected \\
\hline
GX 13+1 & atoll/Z ? & 23 & Quite stable (see Fig 3) & Yes & Radio properties like Z sources \\
2S 0921-630 & ? & & Not in f.o.v. & No & Unlikely to be a Z source \\
\hline
4U 1820-30 & atoll & 12 & Mean hardness$^2$ $\sim 0.78$ & No & Reported detection at $\sim 0.5$mJy level \\
GX 3+1 & atoll & 17 & B (see Fig 3) & No & Improved radio limit \\
GX 9+1 & atoll & 16 & UB (Fig 3) & No &  Improved radio limit \\
GX 9+9 & atoll & 16 & UB? (Fig 3) & No &  Improved radio limit \\
4U 1636-53 & atoll & 12 & Not in f.o.v. & No & First reported radio
limit \\
4U 1705-44 & atoll & 8 & Mean hardness $\sim 0.51$ & No & First reported radio
limit \\
4U 1735-44 & atoll & 9 & Mean hardness $\sim 0.43$ & No & Improved radio limit \\
\hline
GX 339-4 & BHC & 20 & & No & Radio quenched in X-ray High State$^3$\\
\hline
\end{tabular}
\caption{Comparison of radio and X-ray properties of the X-ray
binaries at the time of observation.
($^1$ Z sources : Horizontal Branch (HB), Normal Branch (NB), 
atoll sources : Banana (B), Upper Banana (UB).)
($^2$ Hardness $\equiv$ (5-20 / 2-5) keV count rate) 
($^3$ see Fender et al. 1999b)}
\end{table*}

\subsection{XTE ASM}

The Rossi X-ray Timing Explorer (RXTE) All-Sky Monitor (ASM) monitors
bright X-ray sources up to several times daily in the 2-12 keV
band. In this paper we have utilised the total 2-12 keV intensity data
to judge the state of the X-ray sources on timescales longer than the
relatively short ($\sim 1$ day) SAX WFC observations.  See e.g. Levine
et al. (1996) for more details.  

\section{Discussion}

\subsection{Radio detections and upper limits}

Table 2 summarises the classification (Z, atoll or BHC) of the
targets, and their X-ray and radio state at the time of the
observations. It is clear from the XTE ASM count rates that the Z
sources are systematically brighter than the atoll sources, and that
the `hybrid' source, GX 13+1, is somewhere between the two
groups. Below we discuss each of the source classes in general, as
well as the X-ray : radio connection and the accurate relative
astrometry available from our data.

\subsubsection{The Z sources}

Two of the four Z sources have been detected, as well as the unusual
`hybrid' atoll/Z source GX 13+1. This is not suprising as all of these
sources have been previously detected (e.g. Hjellming \& Han 1995).
We note that these are the same three sources detected in a previous
survey of the same region published by Grindlay \& Seaquist (1986),
although for the non-detections our upper limits are in general
significantly more stringent.  Fig. 1 displays the radio spectra of the
three detected sources, and Table 3 lists the spectral indices
from these data. It is reassuring to note that for GX 17+2, the only
source clearly detected with all baselines {\em and} using only
baselines $\geq 3$km, while the flux densities measured vary by $\sim
50$\%, the spectral index remains about the same. The spectral indices
of GX 17+2 and GX 5-1 are inverted, probably as a result of the
superposition of ejected components which peak earliest at higher
frequencies as they expand. Some variant of the very flat radio--mm
spectral component observed from Cyg X-1 (Fender et al. 2000) cannot
however be ruled out.

\begin{table}
\begin{tabular}{ccc}
Source & \multicolumn{2}{c}{Spectral index $\alpha=\Delta \log S_{\nu}
/ \Delta \log \nu$} \\
       & (all baselines) & (baselines $\geq 3$km) \\
\hline
GX 17+2 & $1.3 \pm 0.1$ & $+1.4 \pm 0.1$ \\
GX 5-1  & --            & $+0.7 \pm 0.1$ \\
GX 13+1 & --            & $-0.4 \pm 0.1$ \\
\hline
\end{tabular}
\caption{Spectral indices for the three sources detected.}
\end{table}

The lack of detection of GX 349+2 and GX 340+0, given their previous
reliable detections (Hjellming \& Han 1995 and references therein),
may be due to our having unluckily caught them in radio `low'
states. This is presumably related to their not having been on the
horizontal branch or upper part of the normal branch in the X-ray HID
- see section 3.2 below.

The upper limits to the linear polarisation of the radio emission from
GX 17+2 and GX 5-1, of order 20\% (Table 1) do not seriously constrain
the emissive mechanism or optical depth of the ejecta. While the
brightest radio transients may show very high linear polarisation
(e.g. Fender et al. 1999a; Hjellming et al. 1999), the total radio flux
(i.e. not resolving individual components) from X-ray binaries is
generally polarised at less than the 20\% level.

\subsubsection{The atoll sources}

Excluding 2S 0921-630 (see discussion below), we have observed several
{\em bona fide} atoll-type X-ray binaries and not detected any of
them. Significantly, this group includes GX 9+1 and GX 9+9, amongst
the most luminous (in X-rays) of the class. In particular we do {\em
not} detect the atoll source 4U 1820-30 in the globular cluster NGC
6624. All the limits presented in Table 1 are either improvemens on
previous limits (typically by a factor of $\sim 2$) or the first
limits presented for the sources.  These upper limits are further
evidence that the atoll sources as a class are significantly fainter
at radio (as well as X-ray) wavelengths than the Z sources, a point
explored further in Fender \& Hendry (2000).

\subsubsection{GX 13+1}

While originally classified as an atoll source, the bright X-ray
binary GX 13+1 shares several of the properties of the Z sources
(Homan et al. 1998 and references therein). In particular the system
is know to be a relatively bright and persistent radio source,
supported by our strong detection at a level $\geq 1$ mJy. The radio
spectrum from this source, unlike that of GX 17+2 and GX 5-1, was
negative during out observations and consistent with optically thin
synchrotron emission. Combined with the relatively bright state, this
suggests the recent ejection of a radio-emitting component which has
expanded to an optically thin state.

The limits on linearly polarised emission, of 8\% and 12\% at 6.3
and 3.5 cm respectively, are fairly constraining as this level of
linear polarisation is sometimes achieved by brighter radio transients
associated with X-ray binaries, particularly when the emission is
optically thin (as suggested in this case by the spectral index of
$-0.4 \pm 0.1$). Probably there is still a large contribution from a
depolarised core and/or Faraday depolarisation in the ejecta.

\subsubsection{2S 0921-690 : an edge-on Z source ?} 

It has been suggested that 2S 0921-630 (V395 Car) is a Z-source viewed
nearly edge-on (Zwarthoed et al. 1993 and references therein; see also
Shahbaz et al. 1999).  As the Z sources have all been detected at one
time or another as radio sources (Hjellming \& Han 1995 and references
therein), it is expected that 2S 0921-630 should also be a radio
source if it is one of these systems. Zwarthoed et al. (1993) placed
upper limits on the radio flux density from this source of $\sim 0.5$
mJy at 4.8 GHz. Fender \& Hendry (2000) have shown that the mean radio
flux density of the Z sources at cm wavelengths, when on the
Horizontal Branch, is $55 \pm 13 / d^2$ mJy (where $d$ is the distance
to the source in kpc). Our significantly stronger upper limits of
$\leq 0.1$ mJy further constrain the nature of 2S 0921-630. It is
possible, given the variability in the radio emission from the
Z sources, that both ourselves and Zwarthoed et al. (1993) have been
unlucky enough to catch 2S 0921-630 in a radio `off' state
(corresponding to the lower NB or FB).  However, if this is not the
case, the stronger upper limits on the radio flux density imply

\begin{enumerate}
\item{2S 0921-630 is not a Z-source and only produces the much weaker
and more transient radio emission associated with atoll-type sources}
\item{2S 0921-630 is a Z-source, at a distance of $\geq 20$ kpc, or}
\item{Radio emission from Z sources is strongly beamed and
perpendicular to the orbital plane, so that the observed flux from 2S
0921-630 is severely reduced. This would however imply that the other
six Galactic Z sources all have very similar inclinations, and we
consider it unlikely.}
\end{enumerate}

We consider the most likely conclusion of this result to be that 2S
0921-630 is not a Z-source.

\subsubsection{GX 339-4}

This black hole candidate is commonly detected at a level of $\sim 5$
mJy at cm wavelengths when in the Low/Hard X-ray state (Hannikainen et
al. 1998 and references therein). However, at the time of these
observations the source was in the High/Soft X-ray state, and the
radio emission was supressed; see Fender et al. (1999b) for a fuller
discussion.

\subsection{The radio : X-ray relation}

Penninx et al. (1988) showed that for the Z source GX 17+2 there
appeared to be a strong coupling between X-ray state and radio
emission. This was in the sense that radio emission was strongest on
the Horizontal Branch (HB), weak on the Normal Branch (NB) and absent
on the Flaring Branch (FB). This relation was found to be consistent
with simultaneous radio and X-ray observations of the Z sources Cyg
X-2 (Hjellming et al. 1990a) and Sco X-1 (Hjellming et al. 1990b).
The relation did not appear to hold in simultaneous radio and
X-ray observations of the Z source GX 5-1 (Tan et al. 1992) but given
the uncertainty in our knowledge of time delays before onset of the
radio emission (presumably as some component of ejecta expands and
increases in radio flux above our detection threshold) more study is
required to investigate this.

We observed four Z sources simultaneously with ATCA and the BeppoSAX
WFC.  Inspection of Fig 2 and Table 2 shows that the two sources
which were detected, GX 17+2 and GX 5-1, were in the upper NB/HB and
HB respectively. The two Z sources which were not detected, GX 340+0 and GX
349+2, were in the NB and at the NB/FB vertex respectively, at the
time of observation. Thus our observations are in agreement with the
relation observed from GX 17+2, Cyg X-2 and Sco X-1, in which radio
emission is significantly stronger on the HB than on the NB, which
itself is stronger than on the FB.

Fig 3 shows the HIDs for the brightest of the atoll sources at the
time of our observations. GX 9+1 appears to be in the UB; while for GX
3+1 and GX 9+9 the X-ray `state' is less well defined. GX 13+1 was
relatively stable during the course of these observations; it will be
of interest to see if recent simultaneous XTE PCA / VLA observations
(Homan et al. in prep) reveal it to be in a similar region of the HID
when radio-bright.

\subsection{Relative astrometry}

Reynolds et al. (1995) have shown that the coordinates measured with
ATCA are accurate to $\sim 17$ mas per degree of separation of the
target from the phase calibrator, plus some systematic error due to
uncertainty in the absolute position of the calibrator itself. Table
{\ref{postable}} lists the positions derived from point-source fits at
3.5 cm to each of the three detected sources. As all three sources
were observed with the same phase calibrator, J1751-253 (B1748-253)
the relative separations can be measured to a few tens of
milliarcseconds. The uncertainty in the absolute position of J1751-253
is no more than 10 mas (it is a VLA `B' quality calibrator), and so
the relative uncertainties due to ATCA will dominate. It is the
improvement in the accurate coordinates of this calibrator which
allows us to improve on the coordinates of GX 5-1 presented in 
Grindlay \& Seaquist (1986).

\begin{table}
\begin{tabular}{ccccc}
\hline
Name & $\Delta \theta$ & \multicolumn{2}{c}{Coordinates (J2000)} & $\sigma$\\
     & (deg) & RA & Dec & (mas) \\
\hline
J1751-253 & -- & 17:51:51.263 & -25:24:0.063 & $\leq 10$ \\ 
\hline
GX 17+2 & 11.4 & 18:16:01.388 & -14:02:10.425 & $\sim 190$ \\
GX 13+1 & 8.3  & 18:14:31.083 & -17:09:25.859 & $\sim 140$ \\
GX 5-1  & 2.3  & 18:01:08.233 & -25:04:42.044 & $\sim 40$ \\
\hline
\end{tabular}
\caption{
Accurate positions of the three sources detected, based upon
point-source fits at 3.5 cm. The dominant positional uncertainty scales as
17 mas per degree of separation ($\Delta \theta$)
of the target from the phase
calibrator J1751-253. The error in each ordinate is $\sqrt{2}$ of
the total uncertainty.}
\label{postable}
\end{table}

The importance of these relative astrometric measurements is twofold.
Firstly, given the crowded nature of the Galactic centre region at
optical and infrared wavelengths, very accurate coordinates are
required to confidently identify the counterpart to the X-ray and
radio source. This is well illustrated by the recent discussion about
the true infrared counterpart to GX 17+2 (Deutsch et al. 1999). Our
best position for this source is consistent with the weighted J2000
position presented in Deutsch et al. (1999) and supports their
assertion that NP Ser is {\em not} the optical/IR counterpart to the
X-ray source. The infrared counterpart to GX 13+1 is more confidently
determined (Naylor, Charles \& Longmore 1991; Bandyopadhyay et
al. 1999). Our new very accurate coordinates for GX 5-1 have recently
been utilised to finally identify the probable infrared counterpart to
this source (Jonker et al. 2000).

Secondly, there is a large uncertainty in distance to most of
these sources. Future accurate relative astrometry will allow
measurements of relative proper motions of the sources to be made to
see whether they agree with the Galactic rotation curve for the
Galactic centre distance, and/or posess high peculiar velocities as
inferred recently for the unusual X-ray binary Cir X-1 (Johnston,
Fender \& Wu 1999).

\section{Conclusions}

We have demonstrated the usefulness of coordinating X-ray and radio
observations of low-mass X-ray binaries. In the absence of large-scale
programs and/or the next generation of radio telscopes, we believe
that such snapshots of the X-ray and radio state of these systems is
the best way to build up a comprehensive (empirical) understanding of
the X-ray : radio relation and hence the accretion disc : jet coupling
in these systems.

\section*{Acknowledgements}

We thank Bob Sault for useful discussions.
The Australia Telescope is funded by the Commonwealth of Australia for
operation as a National Facility managed by CSIRO. The BeppoSAX
satellite is a joint Italian and Dutch programme.  We acknowledge the
use of quick-look results provided by the ASM/RXTE team.  RPF was
supported during the period of this research by EC Marie Curie
Fellowship ERBFMBICT 972436.


\begin{thebibliography}{}

\bibitem{}
Bandyopadhyay R.M., Shahbaz T., Charles P.A., Naylor T., 1999, MNRAS,
306, 417

\bibitem[]{}
Boella, G., Butler, R.C., Perola, G.C., et al., 1997, A\&A, 122, 299

\bibitem[]{}
Burgess A.M., Hunstead R.W., 1995, Pub. Astron. Soc. Aust., 12, 227

\bibitem[]{}
Cooke B.A., Ponman T.J., 1991, A\&A, 244, 358

\bibitem{}
Deutsch E.W., Margon B., Anderson S.F., Wachter S., Goss W.M., 
1999, ApJ, 524, 406


\bibitem{}
Fender R.P., Bell Burnell S.J., Waltman E.B., 1997, Vistas Astron, 41, 3


\bibitem[]{}
Fender R.P., Garrington S.T., McKay D.J., Muxlow T.W.B., Pooley G.G.,
Spencer R.E., Stirling A.M., Waltman E.B., 1999a, MNRAS, 304, 865

\bibitem[]{}
Fender R.P. et al., 1999b, ApJ, 519, L165

\bibitem[]{}
Fender R.P., Pooley G.G.., Durouchoux P., Tilanus R.P.J., Brocksopp
C., 2000, MNRAS, 312, 853

\bibitem{}
Fender R.P., Hendry M.A., 2000, MNRAS, in press ({\bf astro-ph/0001502})

\bibitem{}
Grindlay J.E., Seaquist E.R., 1986, ApJ, 310, 172

\bibitem[]{}
Hannikainen D.C., Hunstead R.W., Campbell-Wilson D., Sood R.K., 1998,
A\&A, 337, 460

\bibitem[]{}
Heise, J., 1998, in: The Active X-ray Sky; Results from BeppoSAX and
   RXTE, Nucl.\ Ph.\ B, 69, 186

\bibitem{} 
Hjellming, R.~M., Han, X., 1995, Radio properties of
X-ray binaries. In : Lewin, W.~H.~G., van Paradijs, J., van der
Heuvel, E.~P.~J. (Eds.), X-ray binaries, Camridge University Press,
Cambridge, 308--330

\bibitem{}
Hjellming R.M., Han X., Cordova F.A., Hasinger G., 1990a, A\&A, 235, 147

\bibitem{}
Hjellming R.M. et al., 1990b, ApJ, 365, 681

\bibitem[]{}
Hjellming R.M. et al., 1999, ApJ, 514, 383

\bibitem{}
Homan J., van der Klis M., Wijnands R., Vaughan B., Kuulkers E., 1998,
ApJ, 499, L41

\bibitem[]{}
Jager, R., Mels, W.A., Brinkman, A.C., et al., 1997, A\&AS, 125, 557

\bibitem{}
Johnston H.M., Fender R.P., Wu K., 1999, MNRAS, 308, 415

\bibitem[]{}
Jonker P.G., Fender R.P., van der Klis M., Hambly N., MNRAS, in press
({\bf astro-ph/0005289})

\bibitem[]{}
Mart\'\i{} J., Mirabel I. F., Rodr\'\i guez L. F., Chaty S., A\&A, 332, L45

\bibitem{}
Mirabel I.F., Rodr\'\i guez L.F., 1999, ARA\&A, 37, 409

\bibitem{}
Naylor T., Charles P.A., Longmore A.J., 1991, MNRAS, 252, 203

\bibitem{}
Nelson R.F., Spencer R.E., 1988, MNRAS, 234, 1105

\bibitem[]{}
Penninx W., 1989, in Hunt J. and Battrick B. (eds), 23rd ESLAB
symposium on `Two Topics in X-ray Astronomy', Bologna, Italy, ESA
SP-296, p.185

\bibitem{}
Penninx W., Lewin W.H.G., Zijlstra A.A., Mitsuda K., van Paradijs J.,
van der Klis M., 1988, Nature, 336, 146

\bibitem{}
Reynolds J.E., et al., 1995, A\&A, 304, 116

\bibitem{}
Shahbaz T., Kuulkers E., Charles P.A., van der Hooft F., Casares J.,
van Paradijs J., 1999, A\&A, 344, 101

\bibitem[]{}
Sault R.J., Teuben P.J., Wright M.C.H., 1995, In: Astronomical Data
Analysis Software and Systems IV, eds. Shaw R., Payne H.E., Hayes J.J.E.,
ASP conf. ser. 77, 433

\bibitem[]{}
Tan J., Lewin W.H.G., Hejllming R.M., Penninx W., van Paradijs J., van
der Klis M., Mitsuda K., 1992, ApJ, 385, 314

\bibitem{} 
van der Klis M., 1995, In: X-ray binaries, Lewin W.H.G., van Paradijs
J., van den Heuvel E.P.J. (Eds), CUP, p.252

\bibitem{}
van der Klis M., 1999, In: Proceedings of the Third William Fairbank
Meeting, Rome, in press, ({\bf astro-ph/9812395})

\bibitem{}
Zwarthoed G.A.A., Stewart R., Penninx W., van Paradijs J., van der
Klis M., Roy A.L., Amy S.W., 1993, A\&A, 267, 101


\end{thebibliography}
\end{document}